\documentclass[twocolumn,preprintnumbers,amsmath,amssymb,aps,prb,reprint]{revtex4}
\usepackage{graphicx}
\begin{document}

\title{
Nonlinear Transport, Dynamic Ordering, and Clustering for Driven Skyrmions on Random Pinning 
} 
\author{
C. Reichhardt and C. J. O. Reichhardt 
} 
\affiliation{
Theoretical Division and Center for Nonlinear Studies,
Los Alamos National Laboratory, Los Alamos, New Mexico 87545, USA
} 

\date{\today}
\begin{abstract}
Using numerical simulations,
we examine the nonlinear dynamics of skyrmions driven over random pinning.
For weak pinning, the skyrmions depin elastically, retaining sixfold ordering; however, 
at the onset of
motion there is a dip in
the magnitude of the structure factor peaks
due to a decrease in
positional ordering,
indicating that the depinning transition can be detected using the structure
factor
even within the elastic depinning regime.
At higher drives the moving skyrmion lattice regains  
full ordering.
For increasing pinning strength, we find
a transition from elastic to plastic depinning that is
accompanied by a
sharp increase in the depinning threshold
due to the proliferation of topological defects,
similar to the peak 
effect found at the elastic to plastic
depinning transition in superconducting vortex systems.
For strong pinning
and strong Magnus force,
the skyrmions in the moving phase
can form a strongly clustered or phase separated state with
highly modulated skyrmion density,
similar to that recently observed in continuum-based simulations
for strong disorder.
As the Magnus force is decreased, the density phase separated state
crosses over to a dynamically phase separated state with uniform density
but with flow localized in bands of motion,
while in the strongly damped limit,  both types of phase separated states are lost. 
In the strong pinning limit,
we find highly nonlinear
velocity-force curves in the transverse 
and longitudinal directions,
along with distinct regions of negative differential conductivity
in the plastic flow regime.
The negative differential conductivity
is absent in the overdamped limit.  
The Magnus force is responsible for both the negative
differential conductivity and
the clustering effects,
since it causes faster moving skyrmions
to partially rotate around slower moving or pinned skyrmions.   
\end{abstract}
\maketitle

\section{Introduction}

There are a wide variety of systems
that can be effectively described   
as a collection of particles that couple to a randomly disordered substrate \cite{1,2}.
Specific examples
include vortices in type-II superconductors with naturally occurring defects \cite{3,4,5,6}, 
colloidal particles on disordered  
landscapes \cite{16,17,18,19,20},
pattern forming systems on rough surfaces \cite{21,22,23,24},  
Wigner crystals in the presence of charged impurities \cite{25,26,27,28}, 
active matter or self-propelled particles in complex environments \cite{29,30},
fluid flow over disordered surfaces \cite{31,32,33}, 
granular matter flowing over disordered backgrounds \cite{34}, various models of 
sliding friction \cite{35}, dislocation dynamics \cite{36,37},
and geological systems such as plate tectonics \cite{38}.
Under an applied drive, these systems typically exhibit a pinned phase at low drive
that transitions
to a sliding state at higher drives,
and
additional transitions can occur within the sliding state
between different types of flowing phases \cite{1,2}. 
In an elastic depinning transition,
the particles keep their same neighbors,
while during plastic depinning,
the particles exchange neighbors,
leading to  a proliferation of topological defects in the form  of
dislocations \cite{1,2,3,39,40}.
Some systems
can be described
in terms of the depinning of polycrystalline states composed of large ordered
regions separated by mobile grain boundaries
\cite{41},
a process that produces
distinct transport features
compared to
strongly disordered
systems in which local ordering extends only a few lattice spacings or less
\cite{39,41}.
These different types of depinning phenomena 
produce different features in the velocity-force curves,
differential conductivity, velocity fluctuation 
spectra, and global structure of the particles \cite{1,2,3}.  
Plastic depinning can be followed by
a dynamical transition at higher drives
from the disordered moving plastic state into a moving crystal \cite{1,3,5}
or moving smectic state \cite{42,43,44,45,46}, where the
system regains considerable ordering when the high velocity motion of
the particles reduces the effectiveness of the pinning.

In certain systems such as vortices in type-II superconductors,
when the
pinning strength is increased or the elastic
constant of the vortex lattice is reduced,
there can be a transition
from elastic to plastic depinning
accompanied
by a sharp increase in the
critical depinning force,
called the ``peak effect,'' which is also associated with changes in the shape of
the velocity-force curves
\cite{2,3,4,47,48}.
The
velocity typically increases monotonically
with driving force $F_{D}$ when the substrate is disordered;
however, the
velocity-force curve
can be nonlinear and exhibit scaling of the form
$V \propto (F_{D} - F_{c})^\beta$, where $F_c$ is the depinning force and the value
of $\beta$ depends on whether the depinning transition is elastic or plastic
\cite{1,2,39,40}.
If the substrate is periodic,
the velocity-force curves can 
show a decrease in the particle velocity with increasing $F_{D}$,
giving rise to negative differential conductivity \cite{1}; however, for random 
substrates,
negative differential conductivity is generally not observed.   

In 2009, the discovery of skyrmions in chiral magnets
opened the study of a new type of particle-like objects that can form an
elastic lattice.
The initial
neutron scattering measurements \cite{49} revealed sixfold ordering,
indicating that the skyrmions form a triangular lattice similar
to the vortex lattices found in type II superconductors \cite{49}.
A short time later,
hexagonal skyrmion lattices were
directly observed with Lorentz microscopy \cite{50}.
Since then, there has been
an enormous increase in skyrmion studies,
and a variety of new materials have been identified that
are capable of supporting skyrmions,
including some in which the
skyrmions are stable at room temperature \cite{51,52,53,54,55,56}. 
It was shown that skyrmions 
can be set into motion by the application of a current,
leading to depinning transitions that can be 
detected via changes in the transport properties \cite{57,58,59,60,61} 
or through direct imaging \cite{51,53,54,56,62,63,64}.
Skyrmions
show great promise for numerous applications due to their size and mobility,
so understanding  
their collective dynamics
is of key importance for manipulating dense skyrmion arrays
\cite{65}.

Among systems that undergo nonequilibrium dynamical transitions
when driven over quenched disorder,
skyrmions
exhibit a new class of behavior,
since unlike
previously studied systems,
the topological nature of the skyrmions causes their
dynamics
to be dominated by a non-dissipative Magnus force
\cite{58}.
The Magnus force
generates a velocity component
perpendicular to the net force experienced by the skyrmion,
and as a result, the
skyrmions move at an angle, called the skyrmion Hall angle $\theta_{sk}$, to an
applied driving force
\cite{54,57,58,62,63,64,66,67,68,69}.
The Magnus force
also strongly modifies the interactions of the skyrmions with pinning sites or barriers.
For localized pinning, the
Magnus force causes the skyrmion to rotate
around the edge of the pinning site,
greatly reducing the depinning threshold
compared to the overdamped limit.
This effect was argued to be one of the reasons
that low
critical depinning forces
have been observed in certain skyrmion systems \cite{54,57,58,58,59,66,67,68,69}.
For line defects or spatially extended pinning sites,
the skyrmions cannot avoid the pinning by moving around it,
so the effective pinning strength remains large even
when the Magnus force is finite
\cite{70,71,72}.

Although the intrinsic skyrmion Hall angle
$\theta^{\rm int}_{sk}$ is a constant 
determined by the material parameters \cite{41},
the measured $\theta_{sk}$ has a strong drive dependence
in the presence of pinning,
as initially observed in particle-based simulations
of skyrmions moving in random \cite{67,73,74} and periodic 
pinning arrays \cite{75}.
At the depinning threshold,
$\theta_{sk} \approx 0$,
and as the skyrmion velocity increases with increasing drive,
$\theta_{sk}$ increases
until it saturates at $\theta_{sk}=\theta_{sk}^{\rm int}$.
Imaging experiments of driven skyrmions reveal
a pinned regime, a creep regime where flow occurs in jumps giving
$\theta_{sk}=0$,
and a viscous flow regime,
where $\theta_{sk}$ 
increases with increasing $F_{D}$ and saturates
at higher drives \cite{62}.
Other experiments have also shown an
increase in $\theta_{sk}$ with increasing drive amplitude \cite{63,64}.
The drive dependence of $\theta_{sk}$
arises when the Magnus force induces a shift
in the motion of the skyrmions as they pass over the pinning sites
\cite{67,69,73,74,75},
similar to the side jump effect found for an electron scattering
off of magnetic impurities.
The faster the skyrmion
moves, the smaller the shift, as illustrated in particle-based simulations.
Continuum-based simulations of driven 
skyrmions show that in the absence of pinning, $\theta_{sk}$ is constant,
while in the presence of pinning, the
velocity-force curves are nonlinear 
and $\theta_{sk}$ increases
with increasing drive
from zero
up to the intrinsic value
\cite{54,68},
indicating that the particle-based skyrmion simulations successfully
capture many features of skyrmion dynamics
even though they do not
include the internal degrees of freedom of the skyrmions.

Simulations reveal that
a transition occurs from a skyrmion crystal
to a disordered skyrmion glass state 
when the pinning strength increases \cite{67}.
The crystal depins elastically 
while the glass depins plastically,
but at higher drives the glassy state can dynamically
order into a moving skyrmion crystal in a process that is
similar to the dynamical ordering
transitions observed in the depinning and
sliding of vortices \cite{1,3,5,44,45,46,47}, colloids \cite{16,18,19}, 
and Wigner crystals \cite{27}.
The Magnus force
causes the dynamical fluctuations of the moving skyrmions to be much more
isotropic than those found in driven overdamped systems,
favoring the emergence of
an isotropic moving crystal
rather than the
moving smectic state
observed in
overdamped systems \cite{74}.
Continuum-based simulations of skyrmions on random
substrates
show similar dynamical ordering under increasing drive
when the substrate is only moderately strong \cite{76}, and
recent experimental neutron scattering data gives evidence
for the dynamic ordering of driven skyrmions at high drives \cite{77}.

In this work we expand upon our previous 
particle-based studies of skyrmion dynamics in random pinning, and
perform a detailed study of
the skyrmion transport, structure, and dynamics
as we vary the pinning strength, the ratio of skyrmions to pinning sites,
and the ratio of the Magnus force to the damping term.
For weak pinning, the skyrmions form a triangular lattice
that depins elastically into a moving triangular lattice.
Even though there is no structural transition between
the pinned and moving states,
we find a dip in the weight of the structure factor peaks at the
depinning transition due to the deviations of the skyrmions from the
lattice positions,
followed by a sharpening of the peaks with
increasing drive as the effectiveness of the pinning decreases.
For increasing pinning strength,
a transition from a pinned crystal state to a disordered skyrmion glass
occurs that
is accompanied by a sharp increase in the critical depinning force $F_{c}$,
similar to the peak effect phenomenon
found at the transition from elastic to plastic
depinning in superconducting vortex systems \cite{3,6,47,48}.
We also find that
the scaling of $F_{c}$ with the pinning strength $F_{p}$
crosses over from a quadratic form in the elastic depinning regime
to linear scaling in the plastic depinning regime \cite{1,2,4}.
These results suggest that
a peak effect phenomenon could be a general feature in skyrmion systems 
that occurs as a function of magnetic field, temperature, or lattice shearing. 
We observe a change in the scaling of the velocity-force curves
across the transition from elastic to plastic depinning, including a region
of negative differential mobility, and we explain
these effects in terms of
the drive dependence of $\theta_{sk}$. 
When the pinning strength is high or the Magnus force is large,
we observe clustered or density segregated states,
where an effective attraction between the skyrmions leads
to the formation of
dense moving bands
separated by low density regions.
In recent continuum-based simulations \cite{76},
a similar clustered or segregated state
appeared at strong pinning  and was
attributed
to the generation of spin waves by the moving skyrmions.
In our work, the effective attraction is
a result of the Magnus force and occurs when the pinning causes
the skyrmions
to move at different relative velocities,
giving them a tendency
to rotate around each other rather than moving
parallel to each other.
The phase separated states only occur when
both the Magnus force and the pinning are sufficiently strong.
For weaker pinning, we find
a dynamical phase separated state in which the skyrmion density is uniform
but
there is a coexistence of localized bands of motion 
and pinned bands.

The paper is organized as follows.  In Section II we describe the details of our simulation
technique.  In Section III we show a transition from elastic to plastic depinning as a
function of pinning strength that is accompanied by a sudden change in the critical
depinning force, called a peak effect,
and we demonstrate that the nonlinearity in the velocity-force curves
is distinct from that found for overdamped systems.
In Section IV we study the impact of the skyrmion density on the peak effect in the
critical depinning force, describe the appearance of transverse locking in the
elastic depinning regime, and show crossing of the velocity-force curves in samples
on either side of the transition from elastic to plastic depinning.
Section V shows how the dynamic phases change when the ratio of the
Magnus force to the damping term is varied.
The density phase separation and
dynamic phase separation
that appear
for strong pinning and strong Magnus force are described in
Section VI.
We summarize our results in Section VII.

\section{Simulation}
We consider  
a two-dimensional system of size $L \times L$
with periodic boundaries in the $x$ and $y$ directions
containing $N$ rigid skyrmions modeled as point particles.
The dynamics of the skyrmions
is obtained by integrating a modified version 
of the Thiele equation that takes into account
skyrmion-skyrmion interactions, skyrmion-pin interactions,
and an external driving force \cite{19,20,21,31}.
The equation of motion of a skyrmion $i$ is 
\begin{equation} 
\alpha_d {\bf v}_{i} + \alpha_m {\hat z} \times {\bf v}_{i} =
{\bf F}^{ss}_{i} + {\bf F}^{p}_{i} + {\bf F}^{D} ,
\end{equation}
where
${\bf v}_{i} = {d {\bf r}_{i}}/{dt}$
is the skyrmion velocity,
$\alpha_d$ is the damping constant which 
tends to align the skyrmion velocity in the direction of the external forces,
and $\alpha_m$ is the strength of the Magnus term
which tends to align the skyrmion velocity in the direction perpendicular to
the external forces.
When both $\alpha_d$ and $\alpha_m$ 
are finite, the skyrmions
move at an angle
called the intrinsic skyrmion Hall  
angle, $\theta^{\rm int}_{sk} = \tan^{-1}(\alpha_{m}/\alpha_{d})$,   
with respect to an externally applied driving force.
The skyrmion-skyrmion interaction is
a short range repulsive force of
the form 
${\bf F}_{i}^{ss} = \sum^{N}_{j\neq i}K_{1}(r_{ij}){\hat {\bf r}_{ij}}$, 
where $K_{1}$ is the modified Bessel function,
$r_{ij} = |{\bf r}_{i} - {\bf r}_{j}|$ 
is the distance between skyrmion $i$ and skyrmion $j$, and
${\hat {\bf r}}_{ij}=({\bf r}_{i}-{\bf r}_{j})/r_{ij}$
\cite{19,30,31}.   
We place $N_{p}$ pinning sites in random but non-overlapping positions
and model each pinning site
as a parabolic trap of range $r_{p}$ that can exert a maximum pinning
force of $F_{p}$ on a skyrmion 
of the form
$ {\bf F}^{p}_i = \sum_{k=1}^{N_p}(F_{p}r^{(p)}_{ik}/r_{p})\Theta(r_{p}-r^{(p)}_{ik})\hat{\bf r}^{(p)}_{ik}$,
where
$r^{(p)}_{ik}=|{\bf r}_i-{\bf r}^{(p)}_k|$ is the distance between
skyrmion $i$ and pin $k$,
${\hat {\bf r}}^{(p)}_{ik}=({\bf r}_i-{\bf r}^{(p)}_k)/r^{(p)}_{ik}$,
and $\Theta$ is the Heaviside step function.
The skyrmion density is $n_{sk}=N/L^2$, the pinning density
is $n_p=N_p/L^2$, and $L=36$.
The driving force is ${\bf F}^{D} = F_{D}{\hat {\bf x}}$,
and
we measure the average velocity
per skyrmion parallel,
$\langle V_{||}\rangle = N^{-1}\sum^{N}_{i=1} {\bf v}_{i} \cdot {\bf \hat x}$,
and perpendicular,
$\langle V_{\perp}\rangle = N^{-1}\sum^{N}_{i=1} {\bf v}_{i} \cdot {\bf \hat y}$,
to the applied drive.
We compute the drive-dependent skyrmion
Hall angle
$\theta_{sk} = \tan^{-1}(\langle V_{\perp}\rangle/\langle V_{||}\rangle)$,
as well as quantities related to the standard deviation of the skyrmion
velocities in the parallel and perpendicular directions,
$\delta V_{||} = \sqrt{ [\sum_{i}^{N_{sk}} (v_{||}^i)^2 - \langle V_{||}\rangle^2]/N_{sk}}$  and
$\delta V_{\perp} = \sqrt{ [\sum_{i}^{N_{sk}} (v_{\perp}^i)^2 -\langle V_{\perp}\rangle^2]/N_{sk}}$.
We also characterize the dynamics as a function of driving force by measuring the
structure factor
$S({\bf k}) = L^{-2} \sum_{i,j} e^{i {\bf k} \cdot {\bf r}_{ij}(t)}$ and the
average
fraction of sixfold-coordinated particles
$P_6=\sum_{i=1}^{N}\delta(6-z_i)$, where
$z_i$ is the coordination number of skyrmion $i$ obtained from a Voronoi tessellation.

\section{Elastic to Plastic Depinning}

\begin{figure}
\includegraphics[width=\columnwidth]{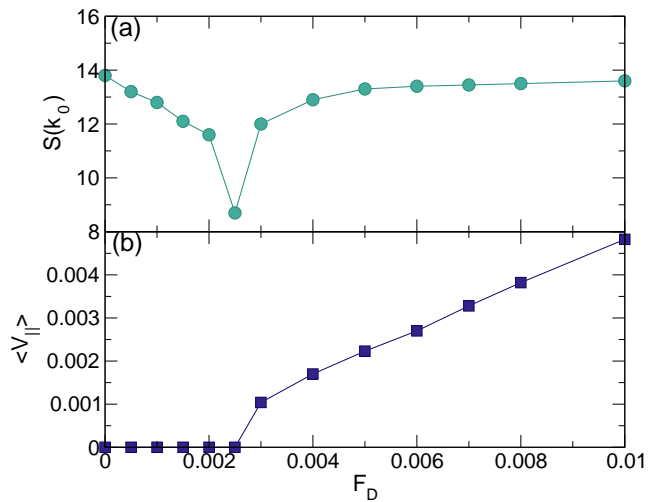}
\caption{
(a) $S({\bf k}_0)$, the magnitude of the structure factor peak at one of the
reciprocal lattice vectors of the skyrmion lattice, vs $F_D$ for
the weak pinning case
$F_p=0.03$ in a sample with $n_{sk}=0.16$, $n_p=0.2$, and $\alpha_m/\alpha_d=9.95$,
where
the skyrmions form a pinned crystal that depins elastically.
(b) The corresponding $\langle V_{||}\rangle$, the average
skyrmion velocity parallel to the driving direction,
vs $F_{D}$.
The dip in $S({\bf k}_0)$ occurs at the drive for which
$\langle V_{||}\rangle$ becomes finite. 
}
\label{fig:1}
\end{figure}

\begin{figure}
\includegraphics[width=0.9\columnwidth]{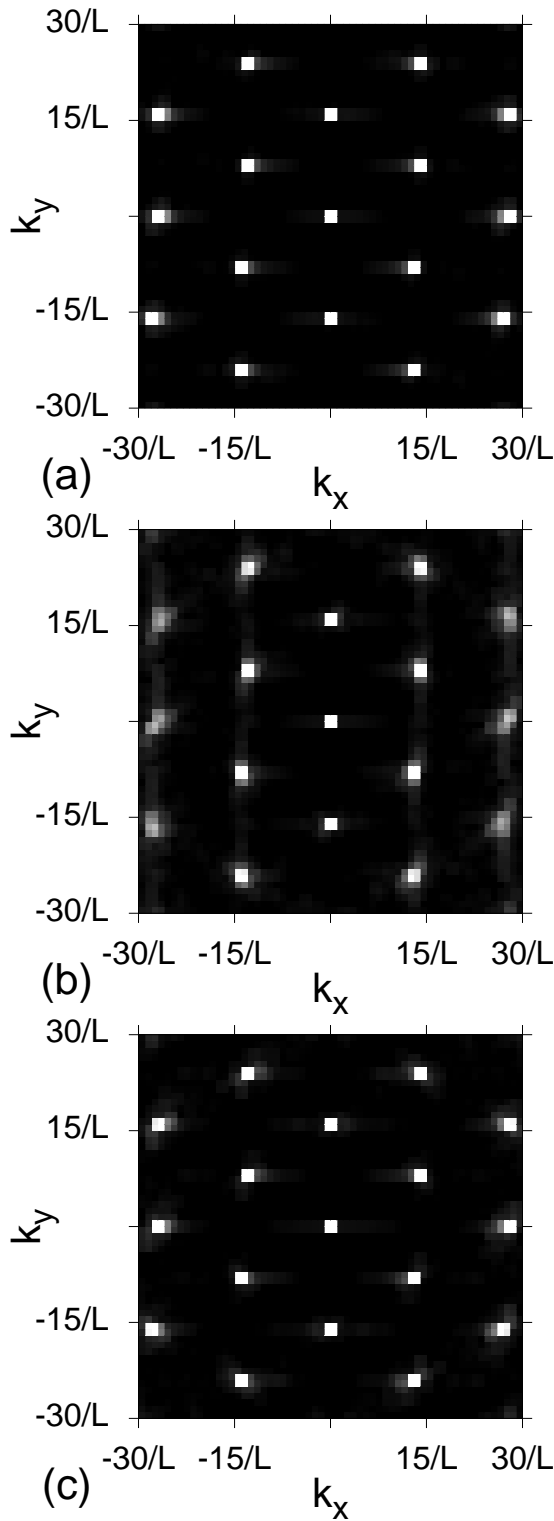}
\caption{ $S({\bf k})$ from the system in Fig.~\ref{fig:1} with $F_p=0.03$,
  $n_{sk}=0.16$, $n_p=0.2$, and $\alpha_m/\alpha_d=9.95$,
  where the depinning is elastic and $F_c=0.0025$.
  (a) At $F_{D} = 0$,
  there are six sharp peaks indicative of triangular ordering.
  (b) Just at depinning for $F_{D}/F_{c} = 1.0$,
  the peaks are still present but there is some smearing.
  (c) At $F_{D} = 0.008$ in the sliding phase, the system is more ordered.  
}
\label{fig:2}
\end{figure}

Previous work with the particle-based skyrmion model showed that
sufficiently strong pinning causes the system to form a glassy
state that depins plastically and then dynamically orders
at higher velocities into a moving
crystal that can be detected using the structure factor $S({\bf k})$ or
$P_6$ \cite{67,74}.
Here we show that even when the pinning is weak, so that
the skyrmions form a pinned lattice rather than a glass and retain their
sixfold ordering across the depinning transition,
there are still detectable changes in the positional ordering of the
skyrmions at the depinning transition.
In Fig.~\ref{fig:1}(a) we plot $S({\bf k}_0)$, the magnitude of the structure
factor at one of the reciprocal lattice vectors ${\bf k}_0$ of the skyrmion
lattice, versus $F_D$ in a sample with
$n_{sk} = 0.16$, $n_{p}  = 0.2$,
$F_{p} = 0.03$,
and $\alpha_{m}/\alpha_{d} = 9.95$,
while in Fig.~\ref{fig:1}(b) we show $\langle V_{||}\rangle$ versus $F_D$ for
the same sample.
Here the skyrmions depin elastically, so
there is no generation of topological defects at the
depinning transition, which occurs at $F_{c} = 0.0025$.
Nevertheless,
there is a dip in $S({\bf k}_0)$ at the depinning transition,
and for $F_D>F_c$, $S({\bf k}_0)$ gradually increases with
increasing $F_D$.
In Fig.~\ref{fig:2}(a) we show $S({\bf k})$
for the same sample
at $F_{D} = 0$ in the pinned state
where there is clear sixfold ordering.
Figure~\ref{fig:2}(b) indicates that
at the depinning transition
$F_{D}/F_{c} = 1.0$,
$S({\bf k})$
maintains its
sixfold ordering
but the peaks become slightly smeared,
while in Fig.~\ref{fig:2}(c)
at $F_{D} = 0.008$ in the moving phase,
the
peaks sharpen again.
We find
a similar trend for other values of $F_{p}$ within the elastic depinning regime. 

\begin{figure}
\includegraphics[width=\columnwidth]{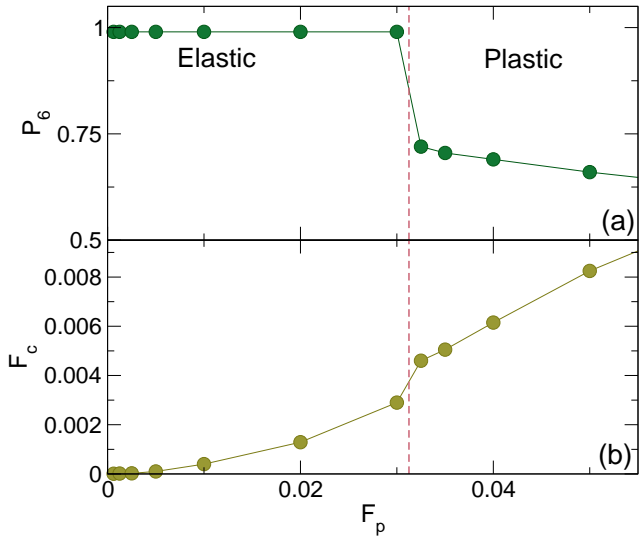}
\caption{
  $P_6$, the fraction of sixfold coordinated skyrmions,
  vs $F_{p}$, measured at the depinning transition for the system in
  Fig.~\ref{fig:1} with $n_{sk}=0.16$, $n_p=0.2$, and $\alpha_m/\alpha_d=9.95$.
  (b) The corresponding value of the critical depinning force $F_{c}$ vs $F_{p}$.
  A transition occurs from elastic depinning,
  where the skyrmions maintain sixfold ordering,
  to plastic depinning,  marked by a drop in $P_{6}$ and a sharp increase in $F_{c}$.
  This
  behavior is similar to the so-called peak effect in type-II superconductors
  that appears across the transition from elastic to plastic vortex depinning.
}
\label{fig:3}
\end{figure}

By conducting a series of simulations and
examining the velocity-force curves and skyrmion positions for
the system in Fig.~\ref{fig:1},
we construct a plot of the fraction $P_6$ of particles with sixfold ordering
at the depinning transition versus $F_p$, shown in Fig.~\ref{fig:3}(a).
For $F_{p} < 0.0325$,
$P_{6} = 1.0$,
since the skyrmions retain six neighbors at
the elastic depinning transition,
while for $F_{p} \geq 0.0325$, 
$P_{6}$ drops
since
there is a
proliferation of topological defects in the form of 5-7 paired dislocations
during plastic depinning.
In Fig.~\ref{fig:3}(b) we plot the
critical depinning force $F_{c}$ versus $F_{p}$. 
For the value of $F_{p}$ at which the drop in $P_{6}$ occurs,
we find a sharp increase in $F_{c}$,
indicating that the skyrmions have become much more strongly pinned.
The sudden increase in $F_{c}$ at the transition from elastic depinning
to plastic depinning
is very similar to the
phenomenon known as the peak effect in type-II superconductors,
where the weakly pinned elastic vortex lattice
undergoes a sharp increase
in the depinning force when the vortex lattice disorders
and becomes well coupled to the pinning \cite{1,3,6}.
In superconducting systems, the strength of the pinning
sites is fixed, but the vortex-vortex interaction strength
can be modified by changing the temperature or
magnetic field, producing
a transition to plastic depinning once the
vortex-vortex
interaction strength
drops below
the vortex-pin interaction strength.
Our results indicate
that a similar peak effect phenomenon     
should be observable in skyrmion systems.

\begin{figure}
\includegraphics[width=\columnwidth]{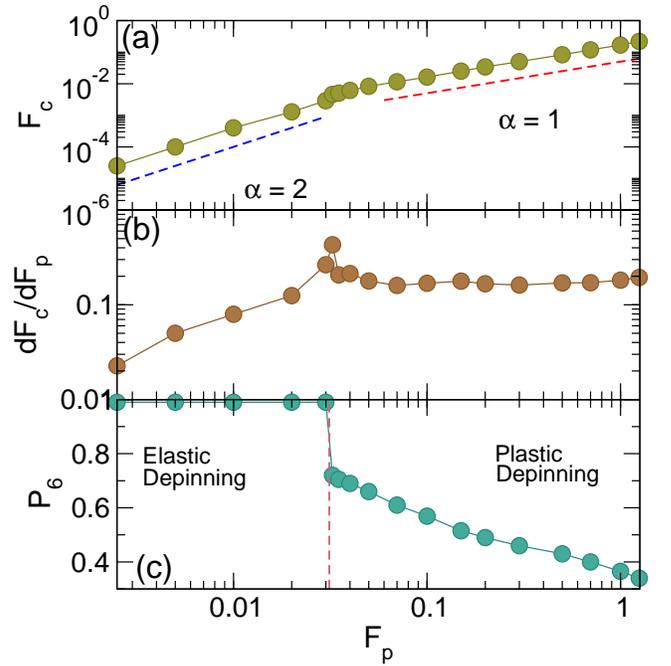}
\caption{(a) $F_{c}$ vs $F_{p}$
  plotted on a log-log scale
for the system in
  Fig.~\ref{fig:1}
  with $n_{sk}=0.16$, $n_p=0.2$, and $\alpha_m/\alpha_d=9.95$.
    The dashed lines are
  power law fits to $F_{c} \propto F_{p}^\alpha$.
  Blue: $\alpha = 2.0$ in the elastic depinning regime;
  red:
  $\alpha = 1.0$ in the plastic depinning regime.
  (b) The corresponding $dF_{c}/dF_{p}$ versus $F_p$ curve has
  a peak near $F_p=0.325$ at the transition from elastic to plastic
  depinning.
  (c) The corresponding $P_{6}$ versus $F_p$ curve shows
  that the elastic to plastic depinning transition coincides with
  a sharp drop in $P_{6}$.   
}
\label{fig:4}
\end{figure}

In Fig.~\ref{fig:4}(a) we plot $F_{c}$ versus $F_{p}$
on a log-log scale
for the system in Fig.~\ref{fig:3},
while in Fig.~\ref{fig:4}(b) we show the corresponding
$dF_{c}/dF_{p}$ versus $F_p$ curve.
In two-dimensional
systems that exhibit elastic depinning, the
depinning force is expected to
scale as $F_{c} \propto F_{p}^\alpha$ with $\alpha = 2.0$,
while in the plastic depinning regime,
there is a similar scaling with $\alpha = 1.0$ \cite{1,4}.
In Fig.~\ref{fig:4}(a), the dashed lines indicate power law fits 
with $\alpha = 2.0$ at smaller $F_{p}$ in the elastic
depinning regime and $\alpha = 1.0$ at larger $F_{p}$ in the plastic
depinning regime,
showing good agreement with
the expected scalings.
At the transition from elastic to plastic depinning, a peak appears in $dF_{c}/dF_{p}$
separating a linear increase of
$dF_{c}/dF_{p}$ with $F_p$ in
the elastic depinning regime
from a constant $dF_{c}/dF_{p}$ in
the plastic depinning  regime, consistent with the power law scaling in the
$F_{c}$ vs $F_{p}$ curves.
The peak in $dF_{c}/dF_{p}$ coincides with a sharp drop in $P_{6}$,
as shown in Fig.~\ref{fig:4}(c),
and $P_{6}$ continues to decrease with increasing $F_{p}$ throughout
the plastic flow regime.

\begin{figure}
\includegraphics[width=\columnwidth]{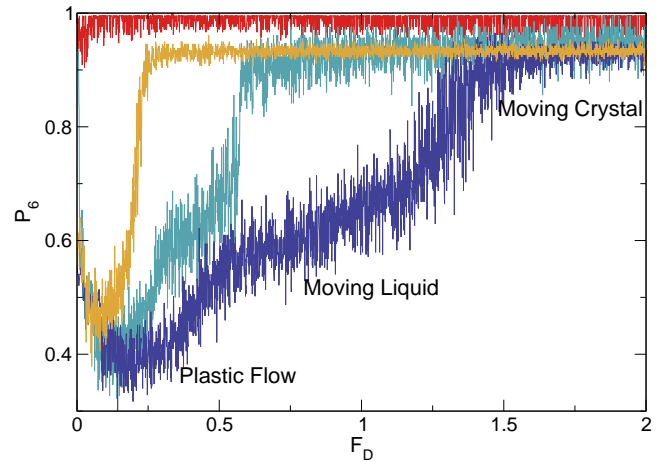}
\caption{$P_6$ vs $F_D$
  for the system in Fig.~\ref{fig:1}
  with $n_{sk}=0.16$, $n_p=0.2$, and $\alpha_m/\alpha_d=9.95$
  at $F_{p} = 0.03$ (red), $0.2$ (orange), 
  $0.3$, (light blue) and $0.5$ (dark blue).
  For large drives, $P_{6}$ increases toward $P_6=1.0$
  as the system dynamically reorders.
  The plateau
  in $P_{6}$
  for $F_p=0.5$ 
  corresponds to a moving liquid phase
  in which all the skyrmions are moving but the system is  
disordered. 
}
\label{fig:5}
\end{figure}

In Fig.~\ref{fig:5} we plot $P_{6}$ versus $F_{D}$
for the system in Fig.~\ref{fig:4}
at $F_{p} = 0.03$, $0.2$,
$0.3$, and $0.5$.
For $F_{p} = 0.03$, $P_{6} \approx 1$
at all values
of $F_{D}$, indicating
that the skyrmions retain their sixfold ordering. 
At $F_{p} = 0.2$ where the depinning is plastic,
$P_{6}$ dips down to $P_6=0.475$ in the plastic flow phase above depinning
but increases
to $P_{6} = 0.93$ near $F_{D} = 0.24$ when the skyrmions
dynamically regain their triangular ordering, similar to what was observed
previously
\cite{67,74}.
For $F_{p} = 0.3$ and $F_p=0.5$,
we find an additional shoulder
feature
above the plastic depinning transition,
and for  $F_{p} = 0.5$ curve this shoulder is followed by
a plateau with $P_6 \approx 0.6$ over the range
$0.5  < F_{D} < 1.4$,
above which the system
dynamically orders.
In the plastic flow phase, 
there is a coexistence
of moving and pinned skyrmions,
while in the moving liquid phase
on the plateau,
all of the skyrmions
are moving but the system is disordered.  

\begin{figure}
\includegraphics[width=\columnwidth]{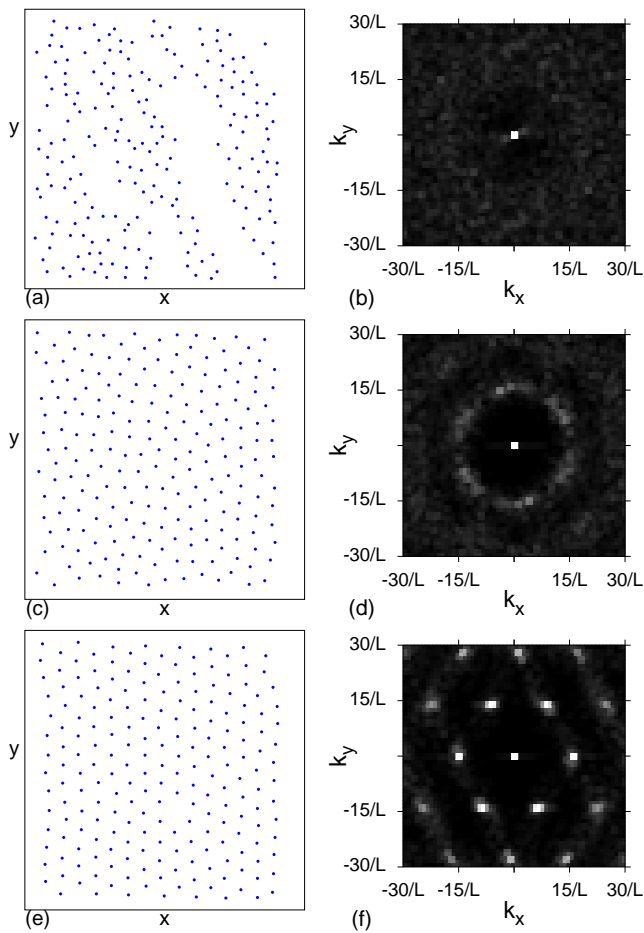}
\caption{(a, c, e) Real space images of the skyrmion positions and
  (b, d, f) the corresponding structure factors $S({\bf k})$
  for the system in Fig.~\ref{fig:5} with $n_{sk}=0.16$, $n_p=0.2$, and
  $\alpha_m/\alpha_d=9.95$
  at $F_{p} = 0.5$. 
  (a,b) At $F_{D} = 0.3$, the plastic flow phase contains both pinned and moving
  skyrmions,
  and $S({\bf k})$ shows that the structure is random.
  (c,d) At $F_{D} = 1.1$, the skyrmion density is more uniform but the
  system is still disordered and forms
  a moving liquid (ML) phase, in which $S({\bf k})$ develops a ringlike feature.  
  (d,e) In the dynamically ordered moving crystal (MC) phase at $F_{D} = 1.75$,
  the skyrmions form a triangular lattice.
}
\label{fig:6}
\end{figure}

In Fig.~\ref{fig:6}(a,b) we show the skyrmion positions and
$S({\bf k})$ for the system in Fig.~\ref{fig:5} at $F_{p} = 0.5$ in the plastic flow phase
at $F_{D} = 0.3$,
where the skyrmion structure is disordered and there is a tendency for
a density phase separation to occur, as discussed in more detail in Section VI.
The structure factor
is nearly featureless,
as expected for a random spatial distribution.
Figure~\ref{fig:6}(c,d) illustrates the skyrmion positions and
$S({\bf k})$ for the same system in
the moving liquid phase at $F_{D} = 1.1$,
where the skyrmion density is more uniform but the skyrmion arrangement is
still disordered, and where the structure factor contains
a ring feature consistent with a liquid state.
We note that it is possible to distinguish between different types of
random structures.
For example, a random arrangement of particles created using a Poisson process
gives a structure factor that has finite weight for $k \rightarrow 0$,
while in an arrangement of particles with what is called a
disordered hyperuniform structure,
$S({\bf k})$ approaches zero as $k \rightarrow 0$
\cite{78}.
It has been argued that in the presence of random pinning, an assembly
of superconducting vortices
exhibits a disordered hyperuniform structure
when the vortex-vortex interactions are
sufficiently strong,
while when the quenched disorder dominates all other energy scales,
a Poisson random arrangement of vortices appears \cite{79}.
One possibility is that the plastic flow phase of the skyrmions is
Poisson random while the moving liquid phase has
a disordered hyperuniform structure.
In Fig.~\ref{fig:6}(e,f) we plot the skyrmion positions and $S({\bf k})$
for the moving crystal phase at $F_{D} = 1.75$, where the skyrmions have
strong triangular ordering and 
$S({\bf k})$ shows
sharp sixfold peaks.

\subsection{Nonlinear Velocity-Force Curves}

\begin{figure}
\includegraphics[width=\columnwidth]{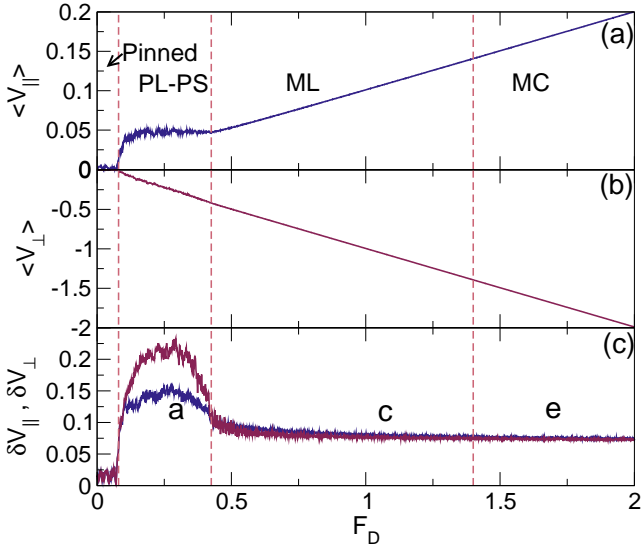}
\caption{(a) $\langle V_{||}\rangle$ and (b) $\langle V_{\perp}\rangle$,
  the average skyrmion velocity parallel and perpendicular to the drive, respectively,
  vs $F_{D}$ for the system in Fig.~\ref{fig:5}
  with $n_{sk}=0.16$, $n_p=0.2$, and $\alpha_m/\alpha_d=9.95$ 
  at $F_{p} = 0.5$. 
Each curve has
different nonlinear behavior near the depinning threshold. 
(c) The corresponding velocity deviations $\delta V_{||}$ (blue) and $\delta V_{\perp}$ (red)
vs $F_{D}$, showing the strong fluctuations in the plastic flow regime.
The letters a, c, and e in panel (c) indicate the values of $F_{D}$
at which the images in Fig.~\ref{fig:6} were obtained.
Vertical dashed lines indicate the separations between the pinned phase,
the plastic flow or phase segregated (PL-PS) state, the moving liquid (ML), and
the moving crystal (MC).
}
\label{fig:7}
\end{figure}

In Fig.~\ref{fig:7}(a,b) we plot
$\langle V_{||}\rangle$ and $\langle V_{\perp}\rangle$ versus
$F_{D}$ for the system in Fig.~\ref{fig:5} at $F_{p} = 0.5$.
Here we find a pinned phase
for $F_{D} < 0.08$,  a plastic flow or phase segregated (PL-PS) phase
for $0.08 \leq F_{D} < 0.425$, a moving liquid (ML) phase
for $0.425 \leq F_{D} < 1.4$, 
and a moving crystal (MC) phase for $F_{D} \geq 1.4$.
In the pinned phase, $\langle V_{||}\rangle = \langle V_{\perp}\rangle = 0$,
and only small shifts in the particle positions occur
as $F_{D}$ is increased.
In the plastic flow or partially phase separated
state, $\langle V_{||}\rangle$ remains nearly constant
while the magnitude of $\langle V_{\perp}\rangle$ increases linearly
with increasing $F_{D}$.
A cusp in $\langle V_{||}\rangle$ appears
at the transition from plastic flow to the moving
liquid phase,
above which $\langle V_{||}\rangle$ begins to increase linearly with $F_D$.
There is  
little change in the behavior of $\langle V_{\perp}\rangle$
through the PL-PS, ML, and MC phases, nor does any
signature of the transition from the ML to the MC phase appear
in $\langle V_{||}\rangle$.
In Fig.~\ref{fig:7}(c) we plot the
velocity deviations $\delta V_{||}$ and $\delta V_{\perp}$ versus $F_D$.
In the PL phase, these velocity fluctuations  
are largest in  the perpendicular direction, while in the ML and MC
phases,
the fluctuations are reduced and become mostly isotropic.
These results show that the strong nonlinearity of
the velocity-force curves
at lower drives
has a very different character than that found
for overdamped particles such as superconducting vortices that exhibit
plastic depinning,
where $\langle V_{||}\rangle$
increases monotonically with $F_D$ according to
$\langle V_{||}\rangle \propto (F_{D} -F_{c})^\beta$ with $\beta \approx 1.5$ \cite{1,2}.
No such scaling can be applied to the
$\langle V_{||}\rangle$ vs $F_{D}$ curve in the skyrmion system,
while the $\langle V_{\perp}\rangle$ curve has $\beta = 1.0$.
Clearly the Magnus
force
strongly modifies
the scaling of the velocity-force curves.
It has been argued that non-dissipative effects
can change the nature of the depinning transition from
continuous to discontinuous
when inertia or stress overshoots are included
\cite{1,80}.
In our system, we find that
the $\langle V_{||}\rangle$ versus
$F_{D}$ curves begin to develop
a discontinuous
jump at depinning
when the Magnus term is finite.

\begin{figure}
\includegraphics[width=\columnwidth]{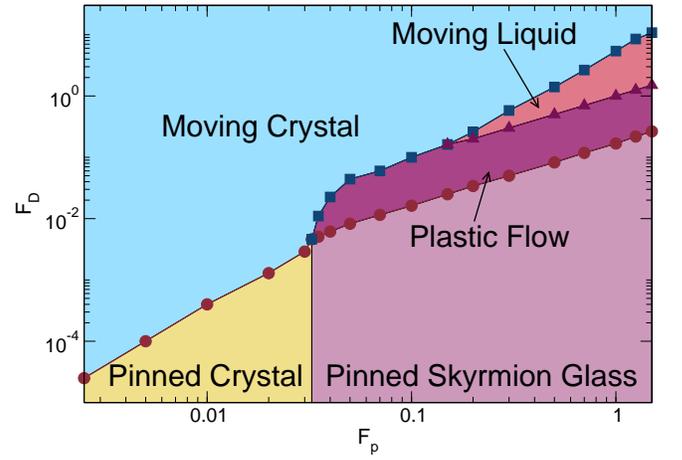}
\caption{Dynamic phase diagram as
  a function of $F_{D}$ vs $F_{p}$ for the system in Figs.~\ref{fig:1}
  to \ref{fig:7} with $n_{sk}=0.16$, $n_p=0.2$, and $\alpha_m/\alpha_d=9.95$,
highlighting the
pinned crystal phase which depins elastically into a moving crystal phase,
the pinned skyrmion glass phase, the plastic flow phase, and the moving liquid phase.    
}
\label{fig:8}
\end{figure}

By conducting a series of simulations for varied $F_{p}$
and analyzing features in the velocity-force and $P_6$ curves,
we
construct the dynamic phase diagram as a function of $F_D$ versus $F_p$
shown in Fig.~\ref{fig:8}.
The pinned crystal phase depins elastically into a moving crystal phase.
In contrast, the
pinned skyrmion 
glass phase depins into a plastic flow phase, which 
transitions into a moving crystal phase for lower $F_p$ or a moving
liquid for higher $F_p$.
The moving liquid phase transitions into a 
moving crystal at higher drives.
For $F_{p} > 0.3$, we find an increasing amount of clustering
occurring in the plastic flow phase.
Overall the phase diagram is similar to
that found in overdamped systems,
except that in the latter, the moving crystal phase
is replaced by a moving smectic phase
since the dynamically generated fluctuations experienced by the particles
are highly anisotropic.
In the skyrmion
system, the Magnus force mixes the fluctuations in the driving direction into
the transverse direction, giving more isotropic fluctuations that cause the
particles to
reorder into a moving crystal rather than a moving smectic state.
The isotropy of the fluctuations increases as the Magnus force
increases, as studied in detail in previous work \cite{74}.

\section{Dynamics as a Function of Skyrmion Density}

\begin{figure}
\includegraphics[width=\columnwidth]{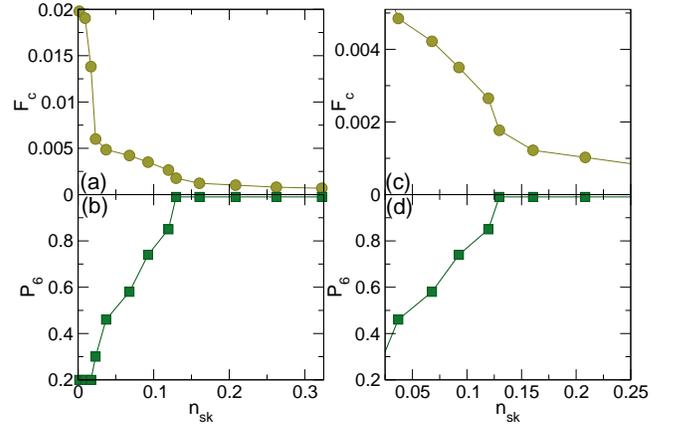}
\caption{ (a) $F_{c}$ vs $n_{sk}$ for samples with
  $n_p=0.2$, $F_{p} = 0.02$,  and $\alpha_{m}/\alpha_{d} = 9.95$.
  (b) The corresponding $P_{6}$ vs $n_{sk}$.
  The transition from elastic to plastic depinning
  occurs near $n_{sk} = 0.125$. 
  For $n_{sk} < 0.02$, the behavior of the system
  is in the single skyrmion limit.   
  (c,d) Zoomed in plots of panels (a) and (b), respectively,
  for the region near $n_{sk}=0.125$ where
  the transition from elastic to plastic
  depinning occurs.
  A drop in $P_{6}$ is correlated
  with an increase in $F_{c}$, which is similar to the peak effect
  phenomenon.
}
\label{fig:9}
\end{figure}

We next consider the effect of varying the skyrmion density while
holding the pinning density and pinning strength fixed.
Such a situation could be achieved experimentally by varying the magnetic
field.
In Fig.~\ref{fig:9}(a,b) we
plot $F_{c}$ versus $n_{sk}$
for samples with $n_p=0.2$, $F_{p} = 0.02$, and $\alpha_m/\alpha_d = 9.95$.
An elastic solid with triangular ordering forms when
$n_{sk} \geq 0.125$, where we find
$P_{6} = 1.0$ and low $F_c$.
For $0.02 \leq n_{sk} < 0.125$,
we observe
a pinned skyrmion glass that depins plastically,
as indicated
by the increase in $F_{c}$ and the
drop in $P_{6}$.
For $n_{sk} < 0.02$, the distance between neighboring skyrmions
becomes so great that the system enters the limit of non-interacting
skyrmions, in which
$F_{c} \approx F_p$
and the skyrmions
are better described as a pinned gas.
Figure~\ref{fig:9}(c,d) shows a blowup of Fig.~\ref{fig:9}(a,b) near
$n_{sk}=0.125$ where the transition from elastic to plastic depinning
occurs.
There is a clear increase in $F_{c}$
at the transition produced when topological defects appear in the system and
$P_6$ drops below 1.
The transition to plastic depinning with decreasing skyrmion density
occurs due to the corresponding decrease in the shear modulus of the skyrmion
lattice.  Once the shear modulus becomes small enough,
topological defects percolate within the lattice, producing a
field-induced peak effect
phenomenon.   

\subsection{Transverse Locking in the Elastic Depinning Regime}

In the elastic depinning regime, the skyrmions depin as a unit and maintain
their triangular lattice ordering.
The moving lattice can become oriented in the direction of drive
or
may remain oriented in some particular direction
due to the geometry of the sample.
In some cases, it may be necessary for topological defects to nucleate inside the
lattice in order to permit it to rotate and orient with the driving direction,
and since these defects are energetically costly, the lattice orientation may
remain fixed in the moving state even when the drive, and the effective
skyrmion Hall angle, increase.

Le Doussal and Giamarchi
argued that a moving superconducting vortex lattice interacting
with pinning and subjected to
a longitudinal drive $F^{||}_{D}$ can exhibit
a finite threshold for transverse depinning
when an additional drive $F^{\perp}_D$
is applied perpendicular to the longitudinal drive, and that 
the lattice can begin sliding along the transverse direction as well
as the longitudinal direction
when $F^{\perp}_{D}$
is large enough
\cite{43}. 
This transverse
critical force has been observed in simulations of moving
superconducting vortices \cite{45,81,82,83,84} and Wigner crystals \cite{27,83} 
as well as in superconducting vortex experiments \cite{85,86}.
Le Doussal and Giamarchi also predicted 
that if the system forms a moving triangular
lattice which remains
oriented in a particular direction, 
then under an additional transverse drive,
the
initial transverse depinning threshold
is followed by
a series of higher-order
transverse depinning thresholds
that appear whenever the vector of net applied force aligns with a symmetry
direction of the moving triangular lattice.
Such directional locking effects were observed in simulations of triangular
vortex lattices moving over random substrates under a fixed longitudinal
drive and an increasing transverse drive \cite{83},
and similar effects
were demonstrated for particles moving over periodic \cite{87,88,89} and
quasiperiodic substrates \cite{91,92} .

\begin{figure}
\includegraphics[width=\columnwidth]{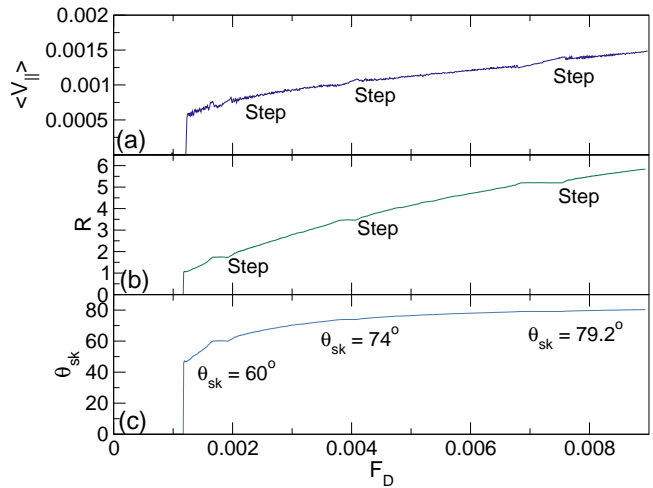}
\caption{The system from Fig.~\ref{fig:9} with $n_p=0.2$, $F_p=0.02$, and
  $\alpha_m/\alpha_d=9.95$
  in the elastic regime at $n_{sk} = 0.207$. (a) 
  $\langle V_{||}\rangle$ vs $F_{D}$.
  (b) $R = \langle V_{\perp}\rangle/\langle V_{||}\rangle$
  vs $F_D$.
  (c) $\theta_{sk} = \tan^{-1}(R)$ vs $F_D$.
  There is
  a series of steps on which the skyrmion lattice motion locks to
  specific skyrmion 
  Hall angles of
  $\theta _{sk} = 60^\circ$, $74^\circ$ and $79.2^\circ$,
  corresponding to the alignment of the direction of motion with a
  symmetry direction of the skyrmion lattice.
}
\label{fig:10}
\end{figure}

In the skyrmion case, we find that
transverse directional locking arises in 
the elastic flow regime
in the {\it absence} of any additional transverse applied
force, which does not occur in the overdamped limit.
The directional locking of the skyrmion lattice
is a result of the velocity dependence of the
skyrmion Hall angle,
which causes the net direction of
motion of the
skyrmion lattice to change as the velocity increases.
When the direction of motion
aligns with a symmetry direction of the
skyrmion lattice,
we find a locking effect,
as shown in Fig.~\ref{fig:10} 
for the system in Fig.~\ref{fig:9} in the elastic regime with
$n_{sk} = 0.207$.
In Fig.~\ref{fig:10}(a) we plot
$\langle  V_{||}\rangle$ versus $F_{D}$
where we observe
three step features.
The $R=\langle V_{\perp}\rangle/\langle V_{||}\rangle$ versus $F_D$ curve in
Fig.~\ref{fig:10}(b)
shows more clearly that 
along these steps, $R$ is constant,
while in the plot of $\theta_{sk}$ versus $F_D$ in
Fig.~\ref{fig:10}(c),
the steps correspond to
skyrmion Hall angles of $60^\circ, 74^\circ$, and $79.2^\circ$.
In each case the skyrmion lattice
remains locked to a specific skyrmion Hall angle 
over a fixed interval of $F_{D}$.
For a triangular lattice,
the
locking
occurs when
$\theta_{sk} = \tan^{-1}(\sqrt{3} p/(2q +1))$,
where $p$ and $q$ are integers.
At $p = 1$ and $q = 0$,
$\theta_{sk}=60^\circ$;
at $p = 2$ and $q = 0$,
$\theta_{sk}=74^\circ$;
and at $p = 3$ and $q = 0$, $\theta_{sk}=79.2^\circ$.
Some smaller steps appear for higher
values of $q$.

\begin{figure}
\includegraphics[width=\columnwidth]{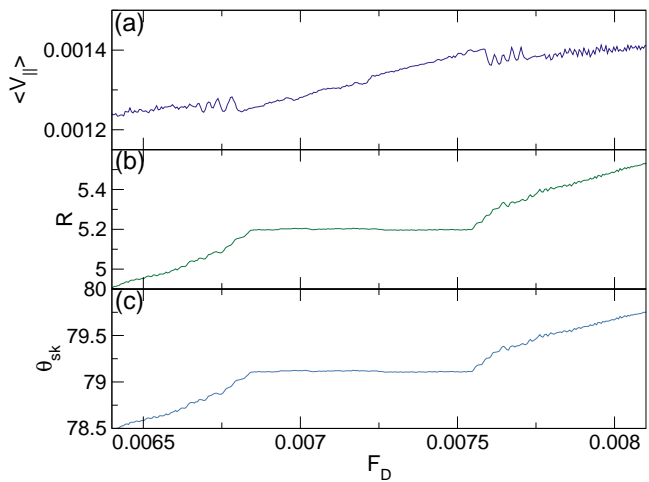}
\caption{A blowup of the
  $\theta_{sk} = 79.2^\circ$
  locking step
  in Fig.~\ref{fig:10}.
  (a) $\langle V_{||}\rangle$ vs $F_D$ has a linear increase along the locking step.
  (b) $R$ vs $F_D$ and (c) $\theta_{sk}$ vs $F_D$ both
  have shapes on either end of the locking step that are consistent
  with what is expected in phase locking systems.  
}
\label{fig:11}
\end{figure}

In Fig.~\ref{fig:11}
we show a blowup of the curves in Fig.~\ref{fig:10} on the
$p =2$, $q = 0$ step
at $\theta_{sk}=79.2^{\circ}$.
Here
$\langle V_{||}\rangle$ increases along the step
while $R$ remains flat
and 
on either side of the step
has the shape
expected
for a
phase locked system such as a Shapiro step.
Figure~\ref{fig:11}(c) shows  that $\theta_{sk}$  is
locked to a fixed angle as well.
We find that the dynamical fluctuations are generally reduced on each step and
that
the peaks in $S({\bf k})$ are sharper,
similar to the enhanced ordering in step regions observed in other systems 
that exhibit phase locking.
These results show that an elastic skyrmion lattice exhibits
a self-induced phase locking that has the same features as phase locking in
an overdamped system but that occurs under the application of a single
fixed direction dc drive
rather than two superimposed dc drives or one rotating dc drive.
In the plastic depinning regime, the dynamically reordered skyrmion lattices
that appear at high drives
generally still contain
a small number of topological defects,
which
smear out the directional locking
features.

\begin{figure}
\includegraphics[width=\columnwidth]{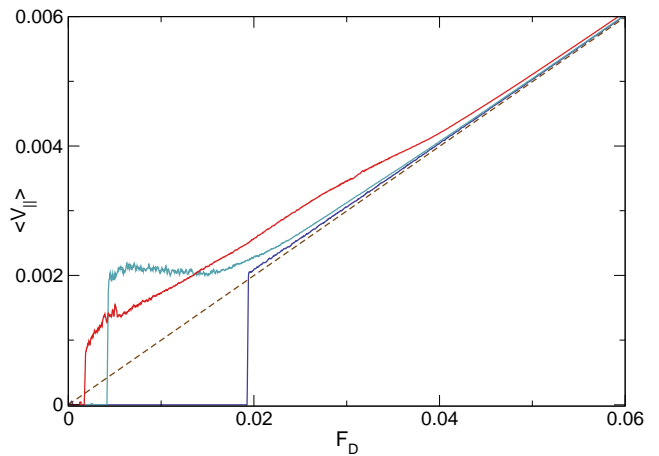}
\caption{ $\langle V_{||}\rangle$ vs $F_{D}$
  for the system in Fig.~\ref{fig:9} with $n_p=0.2$, $F_p=0.02$, and
  $\alpha_m/\alpha_d=9.95$.
  At $n_{sk} = 0.01$ (dark blue), the depinning
  threshold $F_c$ is close to the single particle
  limit and the system does not dynamically order.
  At $n_{sk} = 0.15$ (light blue), the depinning is plastic,
  $F_c$ is higher than in the elastic limit, and
  a plateau appears in the velocity response.
  At $n_{sk} = 0.2$ (red), the depinning is elastic.
  The dashed line indicates the
  velocity response in the pin-free limit,
  showing that the introduction of pinning actually increases
  $\langle V_{||}\rangle$
  due to
  a speed up effect.
} 
\label{fig:12}
\end{figure}

In Fig.~\ref{fig:12} we plot $\langle V_{||}\rangle$ versus $F_{D}$
for the system in Fig.~\ref{fig:9} for varied skyrmion density.
At a low density of $n_{sk} = 0.01$, the system is in the single skyrmion
limit where the critical depinning force $F_{c}$ is high
and the skyrmions do not dynamically order.
Here we find a sharp depinning threshold followed by a monotonic increase
in $\langle V_{||}\rangle$ with $F_D$.
At $n_{sk} = 0.15$, the skyrmions depin plastically and dynamically
reorder at higher drives.
In this case, just above the depinning threshold we find
a window in which $\langle V_{||}\rangle$ remains roughly constant with increasing 
$F_{D}$ in the plastic flow regime before switching to a monotonic increase in
the dynamically reordered regime.
For a higher density of $n_{sk} = 0.2$, $F_c$ is small, the skyrmions
depin elastically, and
$\langle V_{||}\rangle$ increases monotonically
with $F_{D}$.
In the plastic flow window of the $n_{sk}=0.15$ sample
just above depinning, where only some of the
skyrmions have depinned and are moving
while the rest remain pinned, $\langle V_{||}\rangle$
is {\it larger}
than the value found over the same range of drives
for the elastically flowing $n_{sk}=0.2$ sample,
even though in the denser sample, all
of the skyrmions are
moving.
The two velocity-force curves
cross near $F_{D} = 0.175$, above which $\langle V_{||}\rangle$ is higher for
the denser elastically flowing system.
At higher values of $F_{D}$,
$\langle V_{||}\rangle$ converges to the pin-free value (shown as a dashed line) for
all values of $n_{sk}$.
Over a large region of drive, extending from the depinning transition to $F_D=0.04$ and
higher, we find
a pinning-induced speed up effect in which the skyrmions move faster in the direction
of the applied drive than a freely moving overdamped particle, as indicated by the fact
that the velocity-force curves fall above the dashed line.
The pinning-induced speed up effect results when the interaction of the skyrmion
with the pinning site is partially transformed
into motion along the driving direction by the Magnus force.
Such effects have been observed previously
in simulations with
periodic pinning or linelike defects.
For higher drives, the skyrmions move faster, the effectiveness of the pinning
is reduced, and $\langle V_{||}\rangle$ gradually approaches its pin-free value.
If the relative strength $\alpha_m/\alpha_d$ of the
Magnus term
is lowered, the size of the speed up effect
diminishes, 
and in the damping-dominated limit,
$\langle V_{||}\rangle$ is always equal to or less than the pin-free value.
These results show that speed up effects for skyrmions persist even when the
disorder is random rather than periodic.

\begin{figure}
\includegraphics[width=\columnwidth]{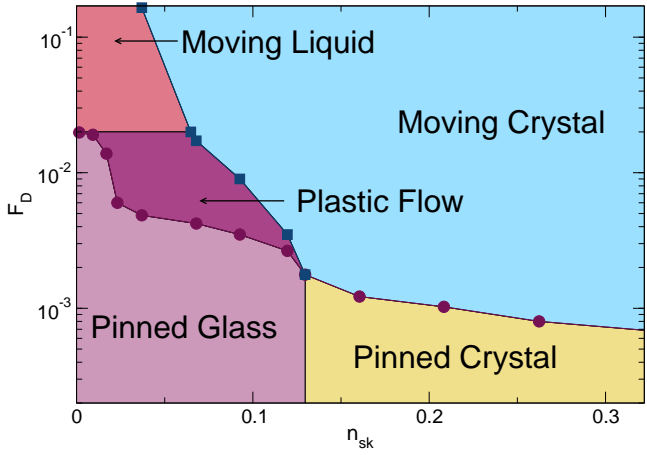}
\caption{ Dynamic phase diagram as a function of $F_{D}$ vs $n_{sk}$
  for the system in Figs.~\ref{fig:9} to \ref{fig:12} with $n_p=0.2$, $F_{p} = 0.02$
  and  $\alpha_{m}/\alpha_{d}=9.95$ highlighting the
pinned glass, pinned crystal, plastic flow, moving liquid, and  moving crystal phases.    
}
\label{fig:13}
\end{figure}

By conducting a seres of simulations, we
construct a dynamic phase diagram as a function of 
$F_D$ versus skyrmion density $n_{sk}$,
as shown in Fig.~\ref{fig:13}.
For $n_{sk} < 0.125$, the system forms a pinned glass, and it may be possible to
further distinguish the formation of a
pinned gas phase for $n_{sk} < 0.02$ where the system enters the single skyrmion limit.
For $n_{sk} \geq 0.125$, the pinned crystal depins into a moving crystal,
while for $0.02 < n_{sk} < 0.125$, the pinned glass undergoes plastic depinning
into a plastic flow phase which transitions into a moving liquid and finally into a
moving crystal at high drives.
For $n_{sk} < 0.02$, above depinning the system 
always remains in a moving liquid phase and dynamical reordering never occurs.  

\section{Dynamics as a function of Magnus Force}

\begin{figure}
\includegraphics[width=\columnwidth]{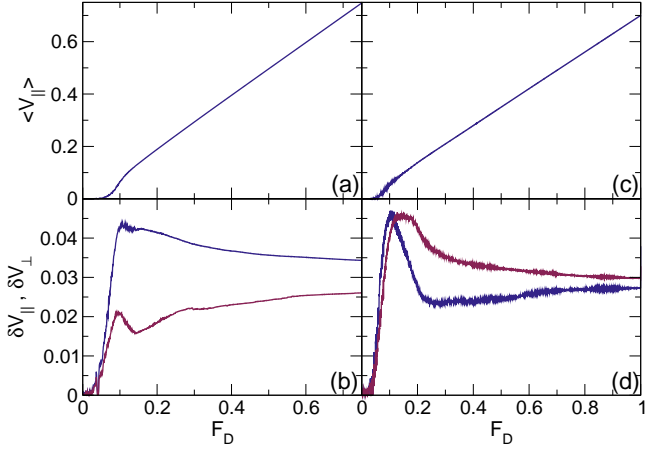}
\caption{ (a) $\langle V_{||}\rangle$ vs $F_D$ and
  (b) $\delta V_{||}$ (blue) and $\delta V_{\perp}$ (red) vs $F_{D}$ for
  a sample with $F_{p} = 0.2$, $n_{sk} = 0.16$,
  and $n_{p} = 0.2$ in the overdamped limit of $\alpha_{m}/\alpha_{d} = 0$. 
  (c) $\langle V_{||}\rangle$ vs $F_D$ and (d)
  $\delta V_{||}$ (blue) and $\delta V_{\perp}$ (red) vs $F_D$ for
  a sample with the same parameters except with 
$\alpha_{m}/\alpha_{d} = 1.0$. 
}
\label{fig:14}
\end{figure}

In Fig.~\ref{fig:14}(a,b) we plot $\langle V_{||}\rangle$, $\delta V_{||}$, 
and $\delta V_{\perp}$
versus $F_{D}$ for a system with
$F_{p} = 0.2$, $n_{sk} = 0.16$, and $n_{p} = 0.2$
in the overdamped limit of $\alpha_{m}/\alpha_{d} = 0$.
Under these conditions, 
$\langle V_{\perp}\rangle = 0$,
and the initial increase
in $\langle V_{||}\rangle$ with $F_D$ just above depinning has the
nonlinear form
$\langle V_{||}\rangle \propto (F_{D} - F_{C})^\beta$ with $\beta \approx 1.3$ to  $1.5$, 
as previously observed for plastic depinning in vortex matter \cite{1,39}.
At higher drives $F_D>0.2$ where all the skyrmions are moving, the
velocity-force curve becomes linear.
We find that in this overdamped system,
$\delta V_{||} > \delta V_{\perp}$ for all drives
since the fluctuating force generated
by the pinning sites is aligned with the driving direction.
The strongly anisotropic pinning-induced fluctuation forces 
cause the system to form a moving smectic,
as predicted from theory \cite{42,43} and observed in
previous simulations \cite{1,45,46,74} and experiments \cite{44}. 
In Fig.~\ref{fig:14}(c,d), we plot $\langle V_{||}\rangle$, $\delta V_{||}$, and
$\delta V_{\perp}$ versus $F_D$ in a system with the same parameters except with
$\alpha_{m}/\alpha_{d} = 1.0$.
Here, the onset of linear behavior in the velocity-force curve shifts closer to the
depinning transition,
while $\delta V_{||} > \delta V_{\perp}$ only in the strongly plastic flow region 
$F_c \leq F_{D} \leq 0.1$.
Within the plastic flow regime,
$\theta_{sk}$ is zero at depinning
and slowly increases toward the intrinsic value $\theta_{sk}^{\rm int}$
with increasing $F_{D}$. 
For
$F_{D} > 0.1$ in the moving crystal phase, we find $\delta V_{\perp} > \delta V_{||}$.
The crossover in the magnitude of the velocity fluctuations occurs
because
the Magnus force induces
fluctuations that  
are perpendicular to the forces exerted by the pinning sites.
At higher drives, $\delta V_{||}/\delta V_{\perp} \rightarrow 1.0$
as the effectiveness of the pinning is reduced.

\begin{figure}
\includegraphics[width=\columnwidth]{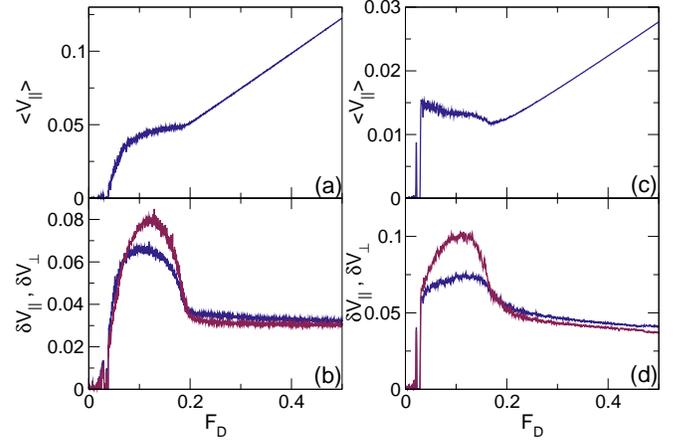}
\caption{
  (a) $\langle V_{||}\rangle$ vs $F_D$  and
  (b) $\delta V_{||}$ (blue) and $\delta V_{||}$ (red) vs $F_{D}$ for 
  the system in Fig.~\ref{fig:14}
  with $F_p=0.2$, $n_{sk}=0.16$, and $n_p=0.2$ 
  at $\alpha_{m}/\alpha_{d} = 4.0$.  
(c) $\langle V_{||}\rangle$ vs $F_D$ and
(d) $\delta V_{||}$ (blue) and $\delta V_{\perp}$ (red) vs $F_D$ for
the same system at
$\alpha_{m}/\alpha_{d} = 15.79$,
where we find a region in which $\langle V_{||}\rangle$
decreases with increasing $F_{D}$,
indicative of
negative differential conductivity. }
\label{fig:15}
\end{figure}

In Fig.~\ref{fig:15}(a,b) we plot $\langle V_{||}\rangle$, $\delta V_{||}$, and
$\delta V_{\perp}$
for the
system in Fig.~\ref{fig:14}
at $\alpha_{m}/\alpha_{d} = 4.0$.
Here the velocity-force curve develops a plateau above depinning,
destroying the scaling associated with plastic flow in the overdamped limit.
For
$F_c < F_D < F_p = 0.2$,
moving and pinned skyrmions 
coexist, and we find
strongly pronounced velocity fluctuations
with 
$\delta V_{\perp} > \delta V_{||}$.
The velocity-force curves become linear
for $F_{D} > 0.2$;
however,
dynamical reordering does not occur until a much higher $F_{D}$, and
the system is in 
a moving liquid phase. 
In Fig.~\ref{fig:15}(c,d) we show $\langle V_{||}\rangle$, $\delta V_{||}$, and $\delta V_{\perp}$ versus $F_D$
for a sample with the same parameters but at
$\alpha_{m}/\alpha_{d} = 15.79$.
Here, there is a region in which $\langle V_{||}\rangle$
decreases with increasing $F_{D}$,
indicative of negative differential conductivity.
For $F_{D} > F_p = 0.2$, 
$\langle V_{||}\rangle$ increases
linearly with $F_{D}$,
while $\delta V_{||}$ and $\delta V_{\perp}$ are
largest for $F_c < F_{D} < 0.19$.
In general, we find that increasing either the Magnus force or $F_p$
increases the range over which
$\langle V_{||}\rangle$ is flat or decreasing with
increasing $F_{D}$, with the maximum range corresponding to $F_c < F_D < F_p$.
Negative differential conductivity,
$d\langle V_{||}\rangle/dF_D < 0$, appears for sufficiently large
Magnus force
in the regime containing
a combination of moving and pinned skyrmions,
and occurs when the increase in $F_D$ combined with the Magnus
force generates an increasing velocity component in the direction perpendicular
to the drive due to collisions between moving and pinned skyrmions,
lowering the velocity response parallel to the drive.
For $F_{D} > F_{p}$, all the skyrmions
depin and the negative differential conductivity is lost.

\begin{figure}
\includegraphics[width=\columnwidth]{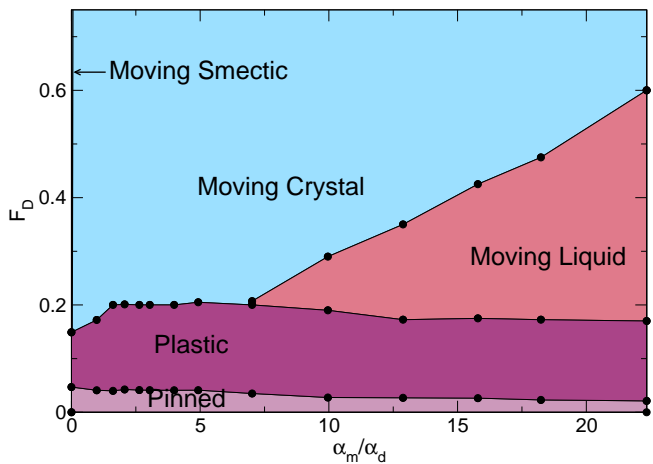}
\caption{ Dynamic phase diagram
  as a function of $F_{D}$ vs $\alpha_{m}/\alpha_{d}$
  for the system in Figs.~\ref{fig:14} and \ref{fig:15} with $F_{p} = 0.2$,
$n_{sk} = 0.16$ and $n_{p} = 0.2$, highlighting
the pinned, plastic, moving liquid, and moving crystal phases.
For $\alpha_{m}/\alpha_{d} = 0$, the
moving crystal phase is
replaced by a moving smectic state.
The depinning threshold $F_{c}$ decreases with increasing 
$\alpha_{m}/\alpha_{d}$.   
}
\label{fig:16}
\end{figure}

By conducting a series of simulations and
examining the nonlinear features in the velocity-force
and $P_6$ versus $F_D$ curves,
we can construct a dynamical phase diagram as a function of
$F_{D}$ versus $\alpha_{m}/\alpha_{d}$,
as shown in Fig.~\ref{fig:16}.
At $\alpha_{m}/\alpha_{d} = 0$,
the pinned and plastic flow phases are followed at higher
$F_D$ by dynamical reordering 
into a moving smectic state,
as described in detail in Ref.~\cite{74}.
For $0 < \alpha_{m}/\alpha_{d} < 6.5$,  
the system dynamically orders into a moving crystal phase
when $F_{d} \approx F_{p}$.
For $\alpha_{m}/\alpha_{d} > 7.5$, we find
a growing window
of a moving liquid state in which
all of the skyrmions are moving  but
there is still considerable topological disorder.
The moving liquid regime increases in extent with larger
$\alpha_m/\alpha_d$ due to the
rotational character of the fluctuations generated by the Magnus force.
If $\alpha_{m}/\alpha_{d}$ is sufficiently large, the Magnus force enhances the
net force experienced by
the skyrmions when
$F_{D}/F_{p} > 1.0$,
and the resulting fluctuations melt the skyrmion lattice.
The magnitude of the fluctuations
grows roughly linearly with $\alpha_{m}/\alpha_d$, while the magnitude of the
forces exerted by the pinning sites decrease as
$1/F_{D}$ \cite{1}, so
the value of $F_{D}$ at which dynamical reordering into the moving crystal
phase occurs
increases roughly       
linearly with $\alpha_{m}/\alpha_d$.
For the illustrated value of $F_{p}$,
the velocity-force curves develop a window in the plastic flow regime
in which $\langle V_{||}\rangle$
remains 
constant or decreases with increasing $F_{D}$.
We note that it is possible for the smectic phase observed at $\alpha_m/\alpha_d=0$ to
persist for finite but small values of $\alpha_{m}/\alpha_{d}$,
as will be discussed elsewhere.
The value of $F_{c}$ marking the depinning transition into the plastic flow state
shifts to lower $F_D$ with increasing $\alpha_{m}/\alpha_{d}$ as
also found in previous simulations,
and
it has been argued that the Magnus force is one
of the reasons that the critical depinning force is small in skyrmion systems
\cite{58}.      

\section{Density Phase Separation and Dynamic Phase Separation}

When both the pinning and the Magnus force are strong, we observe
a dynamically induced density segregation or skyrmion clustering effect. 
This effect was first found in continuum-based simulations of moving skyrmions 
in the strong substrate limit, where it was
attributed to
an attraction between the skyrmions arising from
spin wave excitations generated by the
fluctuating internal modes of the skyrmions
\cite{76}.
Recent simulations with periodic pinning arrays
show that a strong clustering effect occurs for moving skyrmions
when both the Magnus force and the pinning strength
are sufficiently large
\cite{93}.
In these simulations, which contain no spin waves, the effect was attributed to the
drive dependence of the skyrmion Hall angle, which causes different regions of the
skyrmions to move toward each other due to their differing values of $\theta_{sk}$
if a sufficiently strong velocity gradient can be induced by the pinning.
In the case of periodic pinning, this
situation arises both 
in the plastic flow regime
and at higher drives.
Here we show that a similar dynamical density
phase separation
can occur
for strong random pinning,
but that it is restricted to the regime $F_{D}/F_{p} < 1.0$.
We also find two possible types of phase separation: a density modulated state for
strong pinning, and bands of moving skyrmions coexisting with bands of pinned
skyrmions for weaker pinning.

\begin{figure}
\includegraphics[width=\columnwidth]{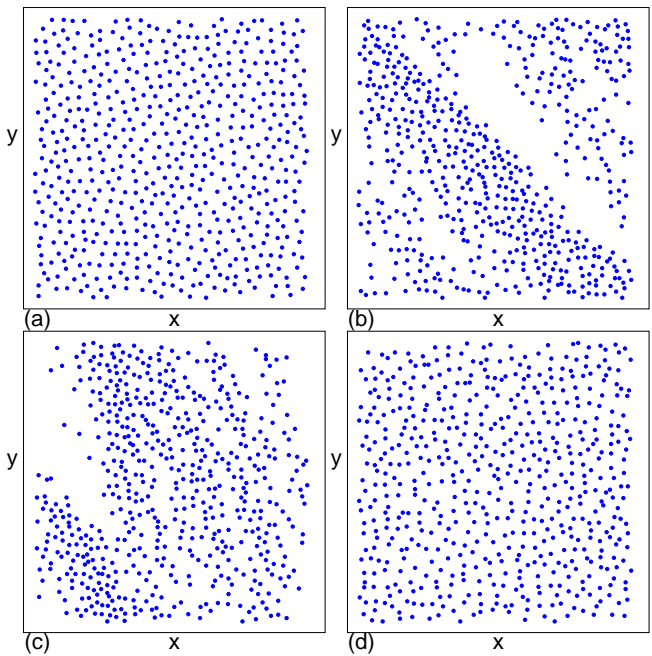}
\caption{Skyrmion positions for a system with 
$F_{p} = 1.5$, $n_{sk} = 0.44$, $n_{p} = 0.6$,  and  $\alpha_{m}/\alpha_{d} = 10$.  
  (a) The pinned phase at $F_{D} = 0.2$ has a uniform
  skyrmion density.
  (b) The density phase separated state at $F_{D} = 0.6$.
  (c) The transition from the density phase separated
  state to the moving liquid state
  at $F_{D} = 1.25$.
  (d) The moving liquid phase at $F_D=2.0$ where the skyrmion density is uniform.
}
\label{fig:17}
\end{figure}

\begin{figure}
\includegraphics[width=\columnwidth]{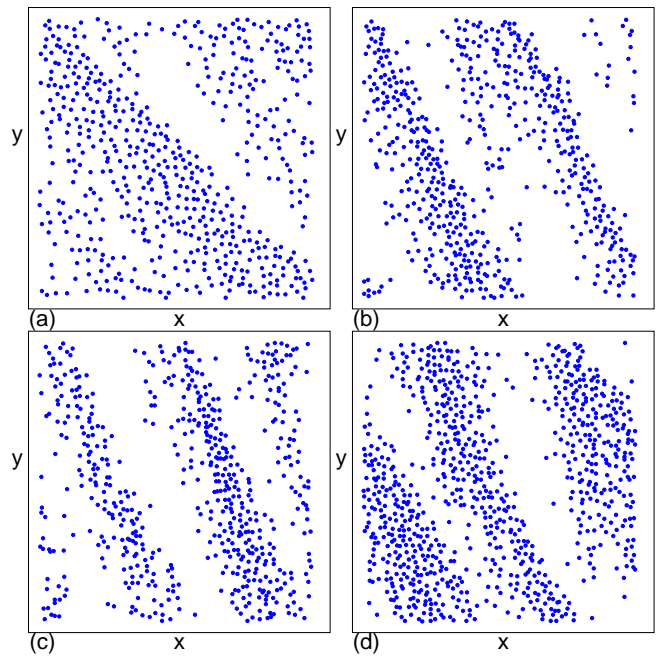}
\caption{
  Skyrmion positions
  in the density segregated state for the
  system in Fig.~\ref{fig:17}
  with $n_{sk}=0.44$, $n_p=0.6$, and $\alpha_m/\alpha_d=10$
  at $F_{D}/F_{c} = 0.6$ for (a) $F_{p} = 1.0$, (b) $F_{p} = 3.5$, and
  (c) $F_{p} = 5.0$, showing that as the disorder strength increases,
  the skyrmion bands become narrower.
  (d) The $F_{p} = 3.5$ system at $n_{sk} = 0.67$, showing the
persistence of the bands at higher skyrmion densities.  
 }
\label{fig:18}
\end{figure}

The density phase
separation is most pronounced
in denser systems,
as illustrated 
in Fig.~\ref{fig:17}
where we show the skyrmion  positions
at different drives for a system with 
$n_{sk} = 0.44$, $n_{p} = 0.6$, $\alpha_{m}/\alpha_{d} = 10$, and  $F_{p} = 1.5$.
In Fig.~\ref{fig:17}(a)
at  $F_{D} = 0.2$, we find a uniform disordered pinned glass state, 
while at $F_{D} = 0.6$ in the plastic flow state,
Fig.~\ref{fig:17}(b) shows that
a dense band of skyrmions emerges that is surrounded by 
a region of low skyrmion density. 
At $F_{D} = 1.25$
in Fig.~\ref{fig:17}(c), near the transition from the density phase separated state
to the moving liquid,
the density  banding is reduced, while in
the moving liquid phase
at $F_{D} = 2.0$ in Fig.~\ref{fig:17}(d), 
the  density phase separation is lost
and the skyrmion density becomes uniform.
In general, the density phase separation occurs
when $F_{p} > 0.5$, $n_{p} > 0.3$, $\alpha_{m}/\alpha_{d} > 5.0$,
and $F_{D}/F_{p} < 1.0$,
in a regime where there is a combination
of moving and pinned skyrmions.
The density phase separation takes the form
of bands of skyrmions roughly aligned with the
direction of motion $\theta_{sk}$,
similar to what was observed
in the continuum-based simulation studies.  
In Fig.~\ref{fig:18}(a)
we show the skyrmion positions
for $F_{D}/F_{p} = 0.6$
in the
system from Fig.~\ref{fig:17}
at $F_{p} = 1.0$, where we find that
the bands of skyrmions  are wider but still present.
At the same value of $F_D/F_c$ for $F_p=3.5$ and 5.0, as
shown in Figs.~\ref{fig:18}(b, c), respectively,
we observe a compression of the skyrmion bands as $F_p$ increases.
Figure~\ref{fig:18}(d) illustrates the
$F_{p} = 0.35$  system
at a higher skyrmion density of $0.67$,
showing that the banding persists for higher skyrmion densities.

\begin{figure}
\includegraphics[width=\columnwidth]{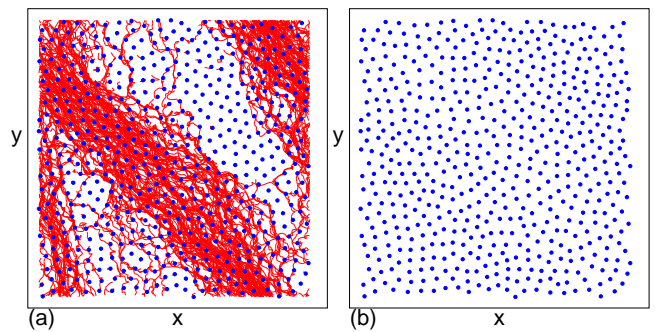}
\caption{ (a) Skyrmion positions (dots) and trajectories (lines)
  for the system in Fig.~\ref{fig:17} with $n_{sk}=0.44$, $n_p=0.6$,
  and $\alpha_m/\alpha_d=10$
  at $F_{p} = 0.3$
  and $F_D=0.1$ where 
  the density phase segregation is lost but
  a dynamical segregation emerges in which
  the motion is confined to a band.
  (b) An image of only the skyrmion positions from panel (a) showing that the
  skyrmion density is uniform.
}
\label{fig:19}
\end{figure}

For the
parameters
illustrated in Figs.~\ref{fig:17} and \ref{fig:18},
we find that
when $0.1 < F_{p} < 0.3$,
the density phase separation disappears
and is replaced by
a dynamical phase segregation in which
the density is uniform but the 
motion occurs only in localized bands.
This is shown in Fig.~\ref{fig:19}(a) where we plot
the skyrmion trajectories
over a period of time in a sample with $F_p=0.3$ at
constant $F_{D} = 0.1$. 
Here
a wide band of moving skyrmions coexists with a pinned band.
Figure~\ref{fig:19}(b) shows only the
particle positions from Fig.~\ref{fig:19}(a) where it is clear that the skyrmion
density is uniform.
For the same parameter set at $0.05 < F_{p} \leq 0.1$,
we find a uniform plastic flow phase,
and for $F_{p} < 0.05$, the depinning is elastic.

\begin{figure}
\includegraphics[width=\columnwidth]{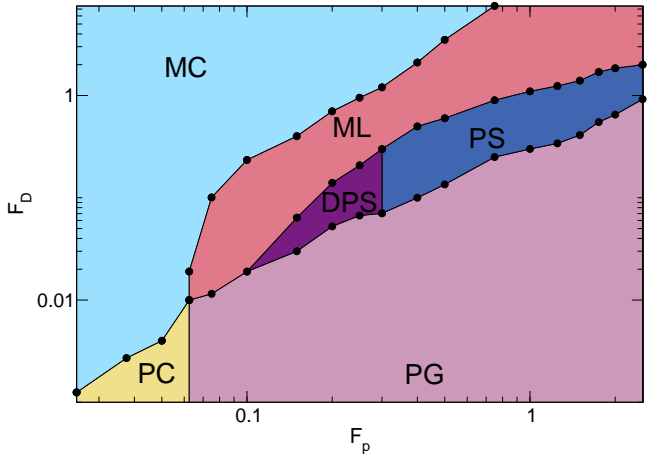}
\caption{Dynamic phase diagram as a function of $F_{D}$ vs $F_{p}$
  for the system in Figs.~\ref{fig:17} to \ref{fig:19} with
  $n_{sk} = 0.44$, $n_{p} = 0.6$, and  $\alpha_{m}/\alpha_{d} = 10$
  showing the pinned crystal (PC),
  the pinned glass (PG),
  the dynamically phase separated state (DPS) illustrated
  in Fig.~\ref{fig:19},
  the density phase separated state (PS) illustrated in Figs.~\ref{fig:17} and \ref{fig:18},
  the moving liquid (ML), and the moving crystal (MC).
}
\label{fig:20a}
\end{figure}

\begin{figure}
\includegraphics[width=\columnwidth]{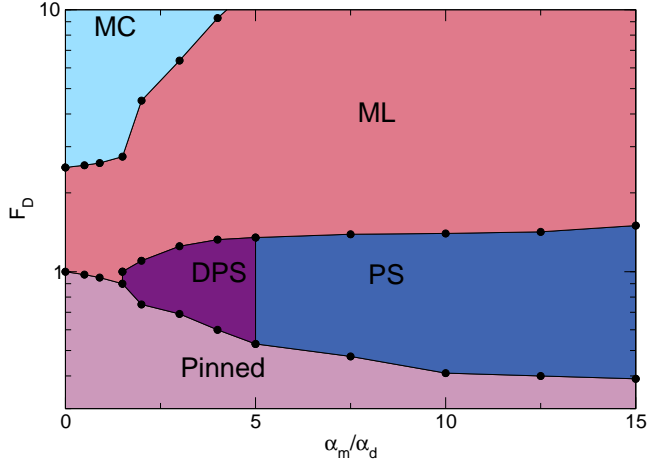}
  \caption{
    Dynamic phase digram as a function of $F_{D}$ vs $\alpha_{m}/\alpha_{d}$
    for the system in Figs.~\ref{fig:17} to \ref{fig:20a} with
    $F_{p} = 1.5$, $n_{sk} = 0.44$, and $n_{p} = 0.6$
    showing the pinned state, dynamically phase separated state (DPS),
    density phase separated state (PS), moving liquid (ML), and moving crystal (MC).
    The PS
    state
    appears only for
    $\alpha_{m}/\alpha_{d} > 5.0$. 
}
\label{fig:20b}
\end{figure}

\begin{figure}
\includegraphics[width=\columnwidth]{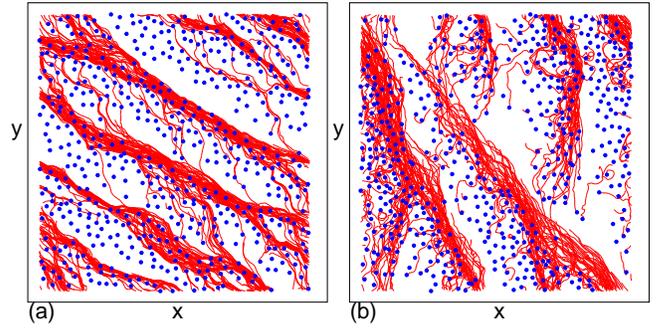}
\caption{Skyrmion positions (dots) and trajectories (lines)
  for the system in Fig.~\ref{fig:20b} with $F_p=1.5$,
  $n_{sk}=0.44$, and $n_p=0.6$
  at $\alpha_{m}/\alpha_{b} = 15$.
  (a) At $F_{D} = 0.3$, multiple bands appear
  consisting of both moving and pinned skyrmions.
  (b)
  At $F_{D} = 1.2$,
  the bands make a steeper angle with the applied drive,
  and kinks appear as the bands rotate to match
  the skyrmion Hall angle, which increases
  with increasing $F_D$.
}
\label{fig:21}
\end{figure}

In Fig.~\ref{fig:20a} we plot a dynamical phase diagram
as a function of $F_{D}$ versus $F_{p}$ for the
system in Figs.~\ref{fig:17} to \ref{fig:19}.
For $F_{p} < 0.05$,
a pinned crystal (PC) phase appears
which depins directly into the moving crystal (MC) phase.
In the moving liquid (ML) phase,
the system is disordered but there is no dynamical or density phase separation.
The dynamically phase separated state (DPS), illustrated in
Fig.~\ref{fig:19},
has uniform skyrmion density but phase separated moving regions,
while the density phase separated state (PS), illustrated in Figs.~\ref{fig:17} and
\ref{fig:18}, has nonuniform skyrmion density.
We find a pinned glass region (PG) at low $F_D$ when $F_p$ is sufficiently large.
In Fig.~\ref{fig:20b} we show the
dynamical phase diagram as a function of $F_D$ versus $\alpha_m/\alpha_d$ for the
same system
at
fixed $F_{p} = 1.5$.
The PS phase appears only when $\alpha_{m}/\alpha_{d} > 5.0$,
while the DPS phase extends from $1.5 < \alpha_{m}/\alpha_{d} \leq 5.0$.
At
$\alpha_{m}/\alpha_{d} = 0$, the MC phase
is replaced by a moving smectic phase (not shown).
When 
$\alpha_{m}/\alpha_{d} > 10$, we find that the bands in the PS phase sharpen and
split into multiple lanes,
as illustrated in Fig.~\ref{fig:21}(a)
at  $\alpha_{m}/\alpha_{d} = 15$ 
and $F_{D} = 0.3$.
Here, multiple bands appear that are each composed of both
moving and pinned
skyrmions.
As $F_{D}$ is increased, the bands simultaneously widen and move at a steeper angle
with respect to the applied drive,
since the skyrmion Hall angle increases with 
increasing drive.
The rotation of the bands as they follow the skyrmion Hall angle
generally occurs via the formation of kinks, in which
one portion of the band
is moving at the lower angle 
while another portion is moving at the new, higher angle,
as shown in Fig.~\ref{fig:21}(b)
at a higher $F_{D} = 1.2$.

\begin{figure}
\includegraphics[width=\columnwidth]{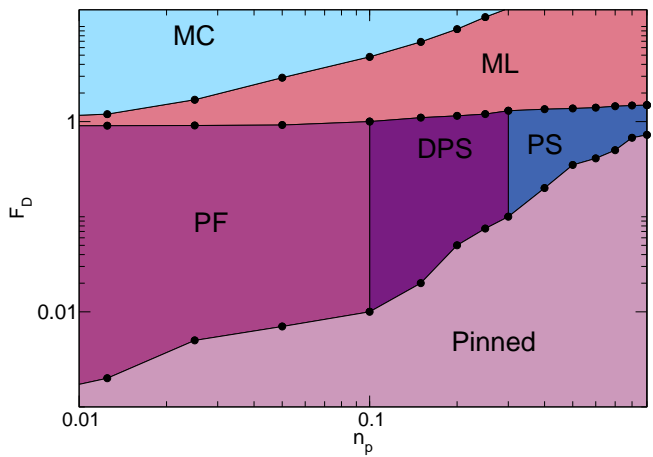}
\caption{ Dynamic phase diagram as a function of $F_{D}$ vs $n_{p}$ for  
a system with $n_{sk} = 0.44$, $F_{p} = 10$, and $\alpha_{m}/\alpha_{d} = 10$
  showing the pinned, plastic flow (PF),
  dynamically phase separated (DPS), phase separated (PS),
  moving liquid (ML), and moving crystal (MC) states.
  In the PF phase, there is a combination of pinned and flowing skyrmions but
  there is no dynamical or density phase separation.
  The DPS and PS states appear only when the disorder is sufficiently dense and
  strong.
}
\label{fig:22}
\end{figure}

In Fig.~\ref{fig:22} we
plot a dynamic
phase diagram as a function of $F_{D}$ versus $n_{p}$
for samples with
$n_{sk}  = 0.44$, $\alpha_m/\alpha_d=10$,  and $F_{p} = 1.5$.
We observe
the PS phase
when $n_{p} > 0.3$ 
and the DPS phase
in the range $ 0.1 < n_{p} \leq 0.3$.
The skyrmion density remains uniform when
$n_{p} < 0.1$ and/or $F_{D} < F_{p}$,
but in this regime
we can distinguish between a
plastic flow (PF) phase
in which moving and pinned skyrmions coexist, and a ML phase in which
all the skyrmions are moving 
but the system is disordered.
The transition to the MC phase shifts to
higher values of $F_{D}$ with increasing $n_{p}$,
while the
transition into the ML phase
remains
roughly constant with increasing $n_{p}$
since it is controlled by the value of $F_{p}$.

The dynamic phase diagrams in Figs.~\ref{fig:20a}, \ref{fig:20b}, and \ref{fig:21}
indicate that a moving segregated state can
arise in particle-based skyrmion models
when both the pinning and the Magnus force are sufficiently strong.
We note that
our particle-based model does not include the generation of spin waves by
the moving skyrmions,
which was the mechanism proposed to be responsible
for the clustering transition observed in continuum simulations. 
The clustering transition we observe arises due to the
velocity differential exhibited by the skyrmions
as they move over the strong random pinning.
The pinned or slowly moving skyrmions 
have a smaller skyrmion Hall angle
than the more rapidly moving skyrmions,
and as a result,
these two types of skyrmions
tend to move toward each other.
It is not clear whether the
clustering effect we observe is actually the same as
that found in the continuum simulations \cite{76} or whether
skyrmion cluster states can be generated via more than one mechanism.
Recent experiments involving
nonequilibrium quenches of a skyrmion system
also show the formation of clustered or phase separated states \cite{94}, so
it is possible that density segregated states are
a general phenomenon in nonequilibrium skyrmion systems.     

\section{Summary} 
Using a particle-based model, we examine
the various types of collective dynamic phases that can occur
for skyrmions moving over random pinning under a dc drive as we vary
the Magnus force, pinning strength, skyrmion
density, and pinning density.
For weak pinning, the skyrmions form a triangular lattice that depins 
elastically, and although there is
no proliferation of topological defects at the elastic depinning transition,
we find that the sixfold peaks in the structure factor diminish in weight 
due to smearing of the skyrmion positions near depinning, while the peaks
sharpen again at higher drives when the skyrmion lattice becomes more ordered.
As a function of increasing pinning strength or decreasing
skyrmion density,
we find
a transition into a pinned skyrmion glass
phase that depins plastically into a disordered moving state.
This transition
is accompanied by a sharp increase in the
critical depinning force
which resembles
the peak effect phenomenon found in superconducting 
vortex systems at the transition from elastic to plastic depinning.

We find that the skyrmion Hall angle is zero at very low drives and increases with
increasing driving force before saturating to the intrinsic value at high drives.
In the elastic depinning regime for weak pinning,
a directional locking effect occurs in which
the direction of skyrmion motion
locks to
a symmetry direction of the triangular pinning substrate.
As the drive increases, the
skyrmion Hall angle also increases,
producing a series of directional locking phases.
These directional locking effects are similar
to those predicted by Le Doussal and Giamarchi
for elastic vortex lattices moving under a longitudinal drive that are subjected to
an additional transverse applied driving force;
however,
in the skyrmion case, the direction of the net driving force remains
fixed and the
skyrmion Hall angle changes due to
interactions with the pinning sites.
We
show that the skyrmion velocity-force curves differ substantially from those
found in overdamped systems with random pinning,
and that when the Magnus force is strong, there can be a plateau or even a
decrease in the skyrmion velocity with increasing driving force,
leading to 
negative differential conductivity.
The scaling of the velocity-force curves changes under strong Magnus force
from the typical form
$V = (F_{D} -F_{c})^\beta$ observed in overdamped systems to a
more discontinuous behavior
similar to that predicted by Schwartz and Fisher for depinning in
systems
that include non-dissipative effects such as
stress overshoots or inertia.
As the Magnus force increases, we find 
a region of what we call a moving liquid phase
in which
the skyrmions are all in motion
but the non-dissipative fluctuations induced by the skyrmion-pin interactions
are large enough to disorder the skyrmion lattice.  Under the same
conditions in the overdamped limit, a moving smectic state appears instead.

One of the most striking features we observe is a density segregation or
skyrmion clustering effect that occurs for strong pinning and strong Magnus force.
Here the skyrmions
form dense bands aligned with the driving direction that are
separated by low density regions, a behavior similar to that recently found
in continuum-based simulations in the strong random substrate limit
as well as in particle-based simulations
of skyrmions driven over periodic pinning arrays.
As the pinning becomes weaker, we find that
the density
phase segregated state
transitions into a dynamically phase separated state in which
the skyrmion density is uniform but
localized bands of moving skyrmions coexist with
bands of pinned skyrmions, while for very weak pinning or small Magnus
force, both the skyrmion density and the skyrmion motion are uniform.
The density segregation arises due to the velocity-dependent skyrmion Hall
angle induced by the Magnus force.
When the pinning is strong enough, the skyrmion velocity can develop a large
local differential,
causing different groups of skyrmions to move at different Hall angles until
the groups collide to form a region of enhanced skyrmion density at the cost of
a density depleted region.
The clustering effects
occur in the plastic flow regime
where there is a combination of pinned and moving skyrmions.
Our results show that a particle-based model that neglects spin waves is sufficient
to capture
such clustering effects;
however, it is
not clear whether the skyrmion clustering
observed in continuum based simulations
is produced by the same mechanism as the clustering found in the
particle-based model.

\begin{acknowledgments}
We gratefully acknowledge the support of the U.S. Department of
Energy through the LANL/LDRD program for this work.
This work was carried out under the auspices of the 
NNSA of the 
U.S. DoE
at 
LANL
under Contract No.
DE-AC52-06NA25396 and through the LANL/LDRD program.
\end{acknowledgments}

\end{document}